# A Universal Strategy of Perovskite Ink-Substrate Interaction to Overcome the Poor Wettability of a Self-Assembled Monolayer for Reproducible Perovskite Solar Cells

*Ashish Kulkarni,\* Ranjini Sarkar, Samah Akel, Maria Häser, Benjamin Klingebiel, Matthias Wuttig, Sudip Chakraborty,\* Michael Saliba,\* Thomas Kirchartz\**

*Ashish Kulkarni, Michael Saliba*
Helmholtz Young Investigator Group FRONTRUNNER, IEK5-Photovoltaik, Forschungszentrum Jülich, Wilhelm-Johnen-Straße, 52428 Jülich, Germany.

*Michael Saliba*
Institute for Photovoltaics (ipv), University of Stuttgart, 70569 Stuttgart, Germany.

*Ranjini Sarkar, Sudip Chakraborty*
Materials Theory for Energy Scavenging (MATES) Lab, Department of Physics, Harish-Chandra Research Institute (HRI), ACI of Homi Bhabha National Institute (HBNI), Chhatnag Road, Jhunsi, Prayagraj 211019, India.

*Ranjini Sarkar*
Ceramic Technologies Group – Center of Excellence in Materials and Manufacturing for Futuristic Mobility, Indian Institute of Technology (IIT) Madras, Chennai, 600036, India.

*Maria Häser*, *Matthias Wuttig*.
I. Physikalisches Institut, RWTH Aachen University, Aachen, Germany.

*Mattias Wuttig*
Peter Grünberg Institute – JARA-Institute Energy-Efficient Information Technology (PGI-10), Forschungszentrum Jülich GmbH, 52425 Jülich, Germany

*Samah Akel, Benjamin Klingebiel, Thomas Kirchartz*
IEK-5 Photovoltaik, Forschungszentrum Jülich, Wilhelm-Johnen-Straße, 52428 Jülich, Germany.

*Thomas Kirchartz*
Faculty of Engineering and CENIDE, University of Duisburg-Essen, Carl-Benz-Str. 199, 47057 Duisburg, Germany.

**Email:**

a.kulkarni@fz-juelich.de (and ashish.kulkarni786@gmail.com)

sudipchakraborty@hri.res.in (and sudiphys@gmail.com)

m.saliba@fz-juelich.de (and michael.saliba@ipv.uni-stuttgart.de)

t.kirchartz@fz-juelich.de

**Abstract**

Perovskite solar cells employing [4-(3,6-Dimethyl-9H-carbazol-9-yl)butyl]phosphonic acid (Me-4PACz) self-assembled monolayer as the hole transport layer has been reported to demonstrate a high device efficiency. However, the poor perovskite wetting on Me-4PACz caused by poor perovskite ink interaction with the underlying Me-4PACz presents significant challenges for fabricating efficient perovskite devices. A triple co-solvent system comprising of dimethylformamide (DMF), dimethyl sulfoxide (DMSO) and N-methyl-2-pyrrolidone (NMP) is employed to improve the perovskite ink – substrate interaction and obtain a uniform perovskite layer. In comparison to DMF and DMSO- based inks, the inclusion of NMP shows considerably higher binding energies of the perovskite ink with Me-4PACz as revealed by density-functional theory calculations. With the optimized triple co-solvent ratio, the perovskite devices deliver high power conversion efficiencies of > 20%, 19.5% and ~18.5% for active areas of 0.16 cm$^2$, 0.72 cm$^2$ and 1.08 cm$^2$, respectively. Importantly, this perovskite ink – substrate interaction approach is universal and helps in obtaining a uniform layer and high photovoltaic device performance for other perovskite compositions such as MAPbI$_3$, FA$_{1-x}$MA$_x$PbI$_{3-y}$Br$_y$ and MA-free FA$_{1-x}$Cs$_x$PbI$_{3-y}$Br$_y$.

## 1. Introduction

The interaction between the perovskite ink and the underlying charge transport layer (CTL) coated substrate strongly determines (1) the uniformity and crystallization of lead halide perovskite thin-films, (2) their charge transport properties,[1-5] (3) radiative and non-radiative recombination processes,[3, 6, 7] (4) reproducibility in the device performance over a large area and (5) the stability.[8-10] The solvents that are used for perovskite ink processing, such as N-N'-dimethyl formamide (DMF) and dimethylsulfoxide (DMSO), are usually highly polar.[11, 12] This enables uniform perovskite layer deposition on hydrophilic CTLs such as $SnO_2$, $NiO_X$, $TiO_2$, achieving high device efficiencies.[13-18] However, using the same perovskite ink to obtain a uniform perovskite film on highly nonpolar and hydrophobic organic hole transport layers (HTLs) such as poly[bis(4-phenyl)(2,4,6-trimethylphenyl)amine] (PTAA) and poly(N,N'-bis-4-butylphenyl-N,N'-bisphenyl)benzidine (poly-TPD) has so far been quite challenging.[19-21] The mismatch between the polarity leads to high contact angles and severe wetting issues of perovskite solution on the hydrophobic HTLs and thus results in poor surface coverage, pinholes, high charge carrier recombination, low open-circuit voltage ($V_{OC}$), and device performance.[10, 22-24] In addition to hydrophobic polymeric HTLs, recently, carbazole-based self-assembled monolayers (SAMs) with phosphonic acid anchoring group such as [2-(9H-Carbazol-9-yl)ethyl]phosphonic acid (2PACz), [2-(3,6-Dimethoxy-9H-carbazol-9-yl)ethyl]phosphonic acid (MeO-2PACz), and [4-(3,6-Dimethyl-9H-carbazol-9-yl)butyl]phosphonic acid (Me-4PACz) have been in the limelight because of their better energy level alignment with some of the most relevant perovskite compositions with high device efficiencies.[25-30] In particular, Me-4PACz-based perovskite solar cells showed one of the highest efficiency for *p-i-n*-type devices both in single junction and tandem (with silicon) configurations.[30] Moreover, in comparison to other SAMs, the Me-4PACz-based device showed the lowest density of interface traps ($2 \times 10^9$ cm$^{-2}$) measured using transient surface photovoltage measurements.[31] This finding implies better extraction of photogenerated charge carriers and higher $V_{OC}$ when the perovskite layer is interfaced with Me-4PACz. Nevertheless, there are only a handful of reports employing Me-4PACz as the HTL in perovskite solar cells[30-35] in comparison to a larger number of reports using 2PACz and MeO-2PACz.[25-29, 36-39] This can be due to the non-uniform perovskite layer coverage on Me-4PACz and low experimental yield.[32, 33, 35] We hypothesize that this is due to the poor perovskite ink (from DMF:DMSO) interaction with Me-4PACz SAM which presents challenges in obtaining a uniform perovskite layer (with full coverage). Previous reports, by Tockhorn et al.,[35] Datta et al.,[32] Farag et al.,[33] and very recent report from Al-Ashouri et al.,[40] on obtaining a non-uniform layer on Me-4PACz from a DMF:DMSO based perovskite ink, also supports our hypothesis. Although engineering of Me-4PACz/perovskite interface by $Al_2O_3$ nanoparticles[34] and deposition of Me-4PACz by evaporation[33] can help to obtain a uniform perovskite layer, the beneficial properties of directly interfacing perovskite with Me-4PACz[31] in the former case and ease of solution process deposition in the latter case might be compromised. Therefore, understanding the reason behind the poor perovskite layer coverage on Me-4PACz and addressing it becomes important. While reports show the use of different SAMs to improve the device performance of perovskite solar cells, there remains little consensus on the fundamental understanding of the perovskite ink interaction with these SAMs, especially with Me-4PACz, as the underlying HTL. On closely comparing the molecular structures of Me-4PACz with PTAA and poly-TPD, as shown in **Figure S1**, one can see the presence of a methyl (or $-CH_3$) and a long alkyl chain ($C_4H_8$), which are in general responsible for the nonpolar and hydrophobic nature.[22, 24, 41-48] This raises similar concerns about issues related to hydrophobicity of Me-4PACz with perovskite inks (from DMF:DMSO), similar to that of PTAA and poly-TPD, and can be the reason for difficulty in obtaining a perovskite layer on, especially, Me-4PACz SAM.

The use of a Lewis acid – Lewis base adduct approach is one of the important strategies to obtain a high-quality perovskite thin film.[49, 50] Various combinations of solvents such as DMF and DMSO,[14] DMF and N-methyl-2-pyrrolidone (NMP),[10, 51] and γ-Butyrolactone (GBL) and DMSO,[12, 52] have been used to dissolve the perovskite precursor materials. In these solvent systems, DMSO and NMP act as Lewis base. Nevertheless, a combination of DMF:DMSO solvent is widely used as DMSO (Lewis base) forms a strong intermediate complex with Pb (II) halides (Lewis acids) owing to its higher donor number compared to the NMP.[53] Although a high-quality perovskite thin film can be achieved with a DMSO-based intermediate complex,[54] obtaining a uniform perovskite layer on highly hydrophobic surfaces presents challenges. Therefore, in addition to the intermediate complex-to-perovskite phase transition via the Lewis acid – Lewis base adduct approach, improving the perovskite ink interaction especially with the underlying HTL (Me-4PACz in present study) without compromising the beneficial properties becomes important to obtain high-quality perovskite solar cells with reproducible performance not only in small area but also in large area devices.

In this work, by fabricating and comparing the $Cs_{0.05}(FA_{0.83}MA_{0.17})_{0.95}PbI_{0.83}Br_{0.17}$ perovskite film formation on Me-4PACz deposited on different indium tin oxide (ITO)/glass substrates we propose a mechanism to explain the challenges associated with obtaining a uniform perovskite layer and achieve a reproducible device performance. Further, to solve the aforementioned issue, we employ a triple co-solvent system composed of DMF, DMSO, and NMP to improve the perovskite ink – Me-4PACz coated substrate interaction and obtain a uniform perovskite layer. Owing to the presence of slightly nonpolar components, the NMP is hypothesized to interact with Me-4PACz. This has been further verified by density functional theory (DFT) calculations showing that binding energies ($\Delta E_b$) and molecular interaction between the perovskite inks and the Me-4PACz coated ITO substrate play a vital role in perovskite film formation. In comparison to DMF-perovskite ($\Delta E_b$ = -4.3 eV) and DMSO-perovskite ($\Delta E_b$ = -4.6 eV) ink systems, the NMP-perovskite ink system shows a considerably higher binding energy ($\Delta E_b$ = -4.75 eV) with the underlying Me-4PACz SAM, thus, resulting in a uniform perovskite layer. In addition to the surface coverage facilitated by the perovskite ink-substrate interaction strategy, the NMP co-solvent ratio (in DMF:DMSO:NMP) is found to play a critical role in determining the device's performance. Triple cation-based perovskite devices obtained with an optimal solvent ratio deliver a champion power conversion efficiency (PCE) of 20 % (area = 0.16 cm$^2$) with negligible hysteresis and high reproducibility. Moreover, the devices demonstrate high stability of $T_{93}$ > ~3600 h.

The poor reproducibility and lack of uniformity of perovskite solar cells make it challenging to obtain a high efficiency with large-area devices. To verify this, with the perovskite-ink – substrate interaction strategy, we successfully fabricated larger-area perovskite solar cells showing efficiencies up to 19.5% and 18.5% in devices with active areas of 0.72 cm$^2$ and 1.08 cm$^2$, respectively. This strategy also shows great universality for fabricating other perovskite components, and devices based on $MAPbI_3$, $FA_{1-x}MA_xPbI_{3-y}Br_y$ and MA-free $FA_{1-x}Cs_xPbI_{3-y}Br_y$ show champion power conversion efficiencies (PCEs) of 19%, 20%, and ~18% respectively, which are competitive PCEs among the reported PCE prepared in the *p-i-n* device structure.

## 2. Results and discussion

*2.1 Hydrophobicity of Me-4PACz*

By employing Me-4PACz SAM as the hole transport layer (HTL) in perovskite (with a bandgap ($E_g$) of 1.68 eV) solar cells, Al-Ashouri *et al.*, reported one of the highest efficiencies in single junction and perovskite-silicon tandem devices.[30] Motivated by these results, we made attempts to fabricate a single junction perovskite solar cell device with Me-4PACz as the HTL. As a perovskite layer we employed the composition $Cs_{0.05}(FA_{0.83}MA_{0.17})_{0.95}Pb(I_{0.83}Br_{0.17})_3$ with an $E_g$ of 1.63 eV. However, a uniform perovskite layer is not formed over the entire Me-4PACz coated substrate which is in line with previous reports.[32, 33, 35, 40] The image of the perovskite layer showing the non-uniform coverage on the ITO/Me-4PACz substrate is shown in **Figure S2a**. To understand this, we measured the water contact angle of Me-4PACz/ITO substrate. As can be seen in **Figure S2b**, Me-4PACz shows a high contact angle of 85° indicating its hydrophobicity. Non-polar groups such as methyl ($–CH_3$) and/or long alkyl chain ($C_4H_8$) present in Me-4PACz (see Figure S1) accounts for its hydrophobic nature. For comparison, the water contact angle on MeO-2PACz is measured. Because of the presence of a comparatively less hydrophobic methoxy ($–OCH_3$) group and a short alkyl chain ($C_2H_4$), MeO-2PACz shows less contact angle. Concomitantly, a uniform perovskite layer is obtained on MeO-2PACz SAM. The image of the perovskite layer on MeO-2PACz and water contact angle measurement are shown in **Figure S2c** and **Figure S2d** respectively. This clearly suggests that methyl ($–CH_3$) and/or long alkyl chain ($C_4H_8$) are responsible for the hydrophobic nature of Me-4PACz SAM.

To further understand this, Me-4PACz is deposited on a fully covered ITO/glass substrate followed by the perovskite layer deposition. Because of the binding of the phosphonic acid group with the ITO over the entire substrate[25] and the existence of π – π interaction between the adjacent carbazole fragments,[55-58] a vertical assembly of very thin (~ 1 to 2 nm) Me-4PACz SAM is formed on the ITO surface.[25] As a result of this, the $–CH_3$ functional group present in Me-4PACz tends to face upward towards the perovskite layer as schematically shown in **Figure 1a**. As $–CH_3$ is non-polar and hydrophobic in nature,[44] initially we hypothesized that this upward facing of the $–CH_3$ is responsible for a non-uniform coverage of the perovskite layer. However, and to our surprise, we obtained a uniform perovskite layer (**Figure 1a**) which could be attributed to a very thin (1 or 2 nm) and uniform vertical assembly of Me-4PACz SAM. On the other hand, when the same Me-

4PACz is deposited on the patterned (or etched) ITO substrate, the perovskite layer is only formed on the ITO part (**Figure 1b**) and not on the glass (without ITO) part of the substrate, which is in line with previous report.[33] Considering the thickness of ITO (~ 100 nm) and binding of the phosphonic acid group with the ITO,[25] a horizontal self-assembly of Me-4PACz SAM on the vertical sides of the patterned (or etched) ITO, as schematically shown in **Figure 1b**, can be envisaged. This could result in the formation of a dense monolayer elongated by the long alkyl chain ($C_4H_8$) of the Me-4PACz, especially on the glass part of the patterned ITO substrate. In addition to this, we suspect a random and dense orientation of Me-4PACz on glass part of the ITO substrate. Previously, Nie et al.,[59] and Nakamura et al.,[60] showed that the use of polar solvents to dissolve SAM results in no SAM layer formation on some oxide surface. In the present study and also in previous reports,[27, 32, 33, 40] ethanol (EtOH), a highly polar solvent, is used to dissolve Me-4PACz. From this it indicates that the use of polar solvent might be the reason for poor/no assembly of Me-4PACz SAM on the glass part of the patterned ITO substrate, thus leading to the formation of a dense Me-4PACz layer. This can also be in MeO-2PACz, however, because of the presence of methyl (–$CH_3$) and long alkyl chain ($C_4H_8$), the surface of the Me-4PACz dense layer becomes hydrophobic. This can cause severe wetting issues with the perovskite ink (in DMF:DMSO) and thus, cause a poor interaction between the perovskite ink and the Me-4PACz coated substrate thereby impede the formation of a uniform perovskite layer. To verify this further, we deposited Me-4PACz on a bare glass substrate (without ITO) followed by the perovskite layer deposition and as expected (**Figure S3**) perovskite layer is not formed. A recent report by Farag et al.,[33] who obtained a uniform perovskite layer on Me-4PACz SAM deposited by evaporation (which does not form a dense layer), also supports the present proposed mechanism. To address this, a triple co-solvent system strategy, schematically shown in **Figure 1c**, with varied ratios of DMF:DMSO:NMP = 4:1-x:x (x = 0 to 1) is employed. This significantly improved the perovskite layer coverage on the Me-4PACz SAM deposited on a patterned ITO substrate. The image of the perovskite thin film showing a uniform coverage on Me-4PACz is shown in Figure 1c.

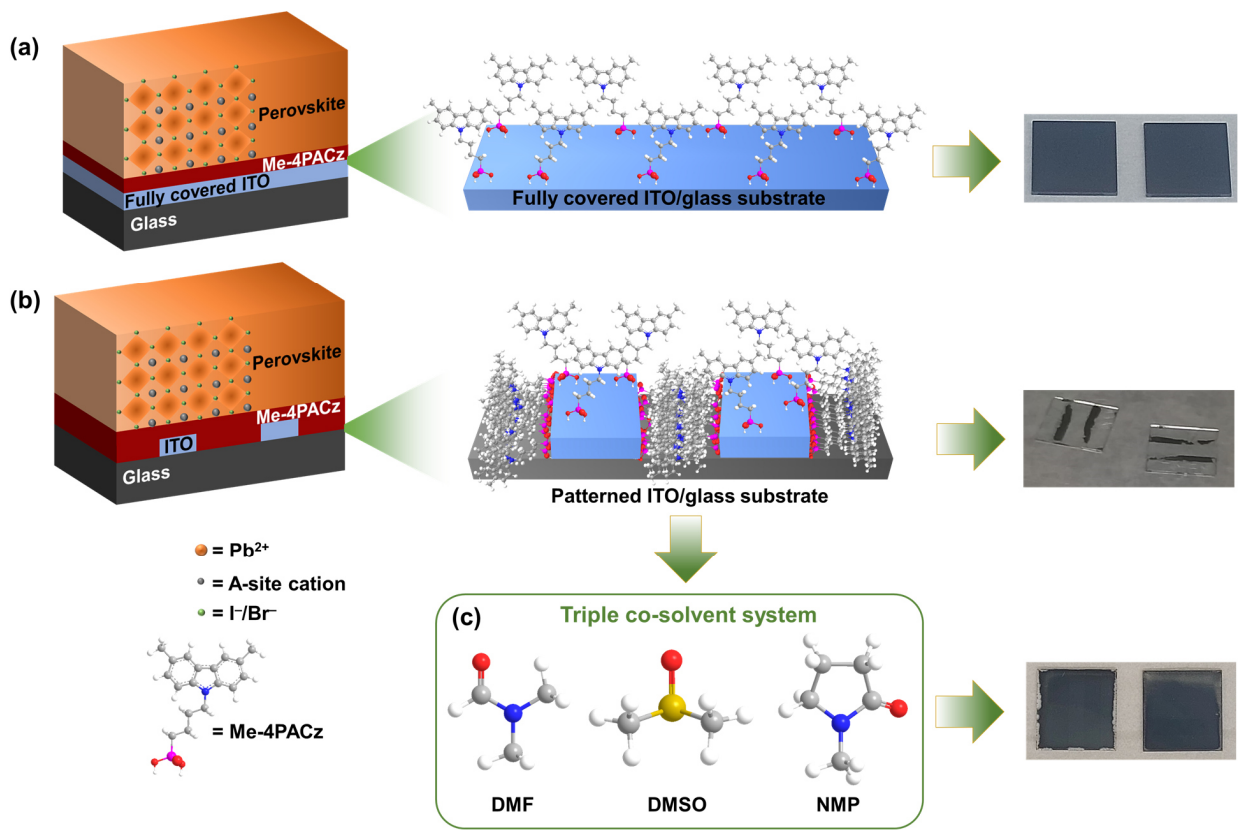

**Figure 1**: Schematic illustration of (a) a uniform vertical self-assembly of Me-4PACz and perovskite layer formation on fully covered ITO/Me-4PACz substrate, (b) horizontal assembly, highly disordered and dense Me-4PACz monolayer and a non-uniform perovskite layer formation on the patterned ITO/Me-4PACz substrate and (c) perovskite ink-substrate interaction strategy by employing triple co-solvent to obtain a uniform perovskite layer on Me-4PACz coated patterned ITO substrate.

*2.2 Triple co-solvent system to improve the perovskite ink – Me-4PACz substrate interaction*

*2.2.1   Importance of NMP for improving the perovskite coverage on Me-4PACz*

In order to understand the formation of a uniform perovskite layer on Me-4PACz SAM, it is important to consider the dipole-related solid-liquid interaction. In this work, the "solid" refers to the Me-4PACz coated ITO substrate and "liquid" refers to the perovskite ink obtained from DMF:DMSO or DMF:DMSO:NMP solvent mixtures. NMP has a lower dipole moment compared to DMF and DMSO (dielectric constants of DMF, DMSO and NMP are summarized in **Table 1**)[61, 62] which makes it a suitable solvent to improve the wetting behavior on hydrophobic surfaces. In addition to this, hydrophilic-lipophilic balance (HLB), defined by $20 \times (1 - (M_L/M)$; where $M_L$ is the lipophilic part and *M* is the overall molecular weight, can also be considered.[63-66] The ability of a polar molecule to interact with non-polar counterparts is determined by a low HLB index. Because of the presence of several non-polar –CH$_2$ groups, the NMP show a low HLB index (see Table 1) when compared to DMF and DMSO.[62] The combination of a low dipole moment and a low HLB for NMP indicates its interaction (through the perovskite ink) with the hydrophobic Me-4PACz coated substrate which could result in a uniform perovskite film.

**Table 1**: Permittivity and HLB index of DMSO, DMF and NMP solvents.[61-66]

| Solvent | Permittivity ($\varepsilon_r$) | HLB index |
|---|---|---|
| **DMSO** | 47 | 12.4 |
| **DMF** | 36.7 | 11.6 |
| **NMP** | 32 | 8.2 |

To understand this further, density functional theory calculations are performed and the binding energies of the solvent-perovskite-Me-4PACz HTL adducts are calculated. The triple cation CsFAMAPbI$_{3-x}$Br$_x$ composition is modelled as the cluster of Cs$^+$PbI$_2$Br$^-$, MA$^+$PbI$_2$Br$^-$, and FA$^+$PbI$_2$Br$^-$ ion pairs (for simplicity, CsFAMAPbI$_2$Br; x=1 is considered). For the explicit description of solvent-perovskite-Me-4PACz systems, each of the solvents (DMF, DMSO, and NMP) is treated separately, instead of considering them as the solvent mixtures. **Figure S4** represents the molecular electrostatic potential (MEP) plots of the isolated solvent molecules (DMF, DMSO, and NMP), triple cation perovskite ion pairs, and Me-4PACz systems. These MEP plots are crucial to predict the fragmental arrangements within the complex structures or the adducts. The electrophilic parts of a molecule are placed in the vicinity of the nucleophilic regions to obtain the most stable complex structures with the strongest inter-fragment interactions. The red, blue, and green regions indicate the regions with negative, positive, and zero electrostatic potentials (ESP), respectively. Evidently, the delocalization of the positive and negative ESP regions within the perovskite ink and solvent molecules are more prominent as compared to the HTL molecule, resulting in higher dipole moments of isolated solvents (4.24D, 4.44D, and 4.17D for DMF, DMSO, and NMP) and the perovskite cluster (13.12D) compared to Me-4PACz (1.50D). Notably, the red surface around the oxygen atoms within the MEP surfaces of the solvents (Figure S4) indicates that the oxygen atoms are the centers for electrophilic interactions.

We simulated molecular interactions to calculate the binding energies of solvent molecules independently with perovskite cluster and Me-4PACz and perovskite-Me-4PACz as a whole. The MEP plots of the optimized molecular structures of perovskite-solvent, Me-4PACz-solvents and solvent-perovskite-Me-4PACz adducts are depicted in **Figure 2a**, **Figure 2b** and **Figure 2c**, respectively, while the calculated binding energies are summarized in **Table S1**. Importantly, two solvent molecules are considered for each perovskite-solvent adduct, one solvent molecule is considered for each HTL-solvent adduct, and two solvent molecules are considered within each perovskite-Me-4PACz-solvent adduct. Within solvent-Me-4PACz adducts, the O atoms are surrounded by negative ESP (red) surface, indicating electron density localization at these atoms. On the other hand, within solvent-perovskite and solvent-perovskite-Me-4PACz adducts, O atoms indicate positive ESP (blue) regions, signifying electron density delocalization. These observations corroborate stronger inter-fragment interactions within solvent-perovskite and solvent-perovskite-Me-4PACz adducts compared to the solvent-Me-4PACz adducts. The binding energy ($\Delta E_b$) of NMP-perovskite is calculated to be -2.44 eV, which is more negative than that between DMSO-perovskite ($\Delta E_b$ = -2.40 eV) and DMF-perovskite ($\Delta E_b$ = -2.22 eV). This indicates that the NMP interacts strongly with the overall perovskite cluster and is in good agreement with previous reports.[10, 51] The Me-4PACz SAM is found to interact with all the solvent molecules through the formation of a hydrogen bond. However, in comparison to the DMF-Me-4PACz and DMSO-Me-4PACz, the binding energy is found to be more negative for NMP-Me-4PACz (see Table S1) indicating a higher affinity of the NMP solvent towards Me-4PACz SAM. The optimized molecular geometries of NMP-perovskite-Me-4PACz (Figure 2c) show a higher binding energy than DMSO-perovskite-Me-4PACz and DMF-perovskite-Me-4PACz. This trend can be correlated with the experimental observation that the NMP co-solvent interacts with Me-4PACz SAM, especially on the glass part of the patterned ITO substrate, and improves the perovskite layer coverage. To further confirm this conclusion, we performed contact angle measurements of Me-4PACz by using perovskite ink in DMF:DMSO and DMF:DMSO:NMP. The lower contact angle in DMF:DMSO:NMP case (**Figure 2e**) compared to the case of DMF:DMSO (**Figure 2d**) suffices improved interaction of NMP-based perovskite ink with Me-4PACz and the formation of a uniform perovskite layer on Me-4PACz.

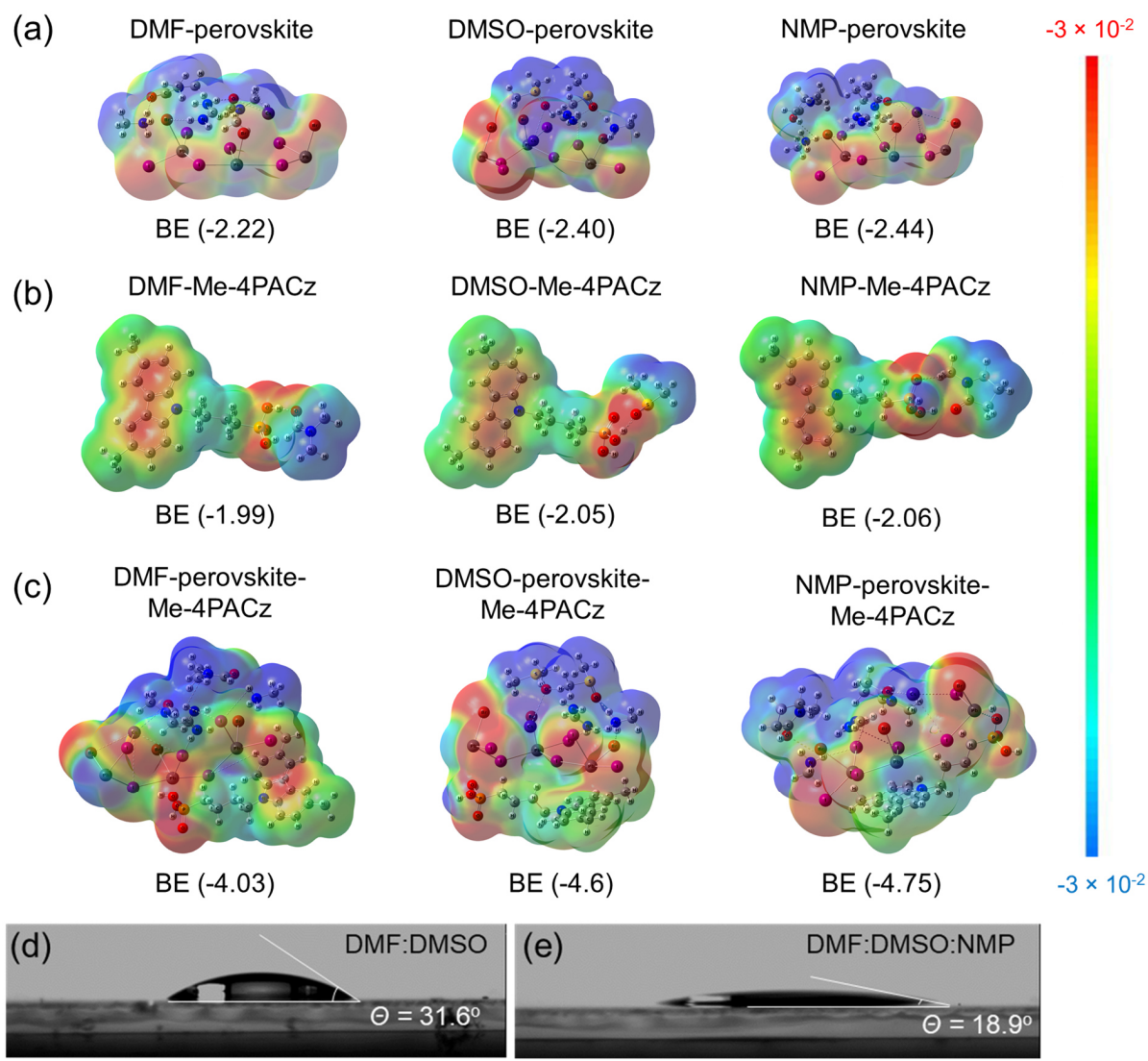

**Figure 2**: Molecular electrostatic potential (MEP) plots of (a) solvent-perovskite adducts, (b) solvent-Me-4PACz adducts, (c) solvent-perovskite-Me-4PACz adducts, and contact angle measurement on Me-4PACz with perovskite ink from (d) DMF:DMSO and (e) DMF:DMSO:NMP solvents.

*2.2.2   Thin film properties*

Perovskite thin films are prepared on Me-4PACz coated substrate by spin coating the solution in different DMF:DMSO:NMP (4:1-x:x) solvent ratios followed by antisolvent dripping and annealing (at 100 °C for 30 min), see supporting information for more details. With increasing the amount of NMP in DMF:DMSO:NMP solvent ratio, the perovskite layer coverage on Me-4PACz improved significantly. The photographic images of perovskite thin films are shown in **Figure S5**.

The X-ray diffraction (XRD) data of perovskite thin films obtained from different solvent ratios show the characteristic perovskite peak at 2θ = 14° as shown in **Figure 3a**.[14, 25] From 4:1 (without NMP) to 4:0.4:0.6 solvent ratios, the XRD pattern of perovskite films shows an additional peak at 2θ = ~ 12° corresponding to residual $PbI_2$, as previously reported for triple cation perovskites.[25] Upon further increase in the NMP solvent ratio (4:0.2:0.8 and 4:0:1 (without DMSO)), the residual $PbI_2$ peak disappears (Figure 3a and **Figure 3b**) and a new peak at 2θ = ~8° appears which corresponds to NMP – $PbI_2$ crystalline solvent intermediate complex.[67, 68] The DMSO and NMP are Lewis bases and are known to form a crystalline intermediate complex with $PbI_2$ (Lewis acid) by intercalating into the $[PbI_6]$ octahedral layers.[67, 69, 70] Among DMSO and NMP, DMSO shows a higher tendency to interact with $PbI_2$ and form an intermediate complex owing to its higher donor number (29.8) than NMP (27.3).[53] Therefore, the formation of DMSO – $PbI_2$ intermediate complex is favored in the perovskite film obtained from 4:1 (without NMP) to 4:0.4:0.6 solvent ratios. It is important to note that the solvent (DMSO or NMP) is forming an intermediate complex with the residual $PbI_2$ which is different from the solvent (DMSO or NMP)-$PbI_2$-AI (A = $FA^+$ or $MA^+$) intermediate complexes. During the perovskite crystallization (annealing step), the solvent DMSO evaporates resulting in the appearance of a residual $PbI_2$ peak (Figure 3a). On the other hand, in the case of 4:0.2:0.8 and 4:0:1 (without DMSO) solvent ratios, the amount of DMSO is less/absent and therefore, NMP Lewis base dominates. This results in the formation of an NMP – $PbI_2$ intermediate complex. As NMP has a higher boiling point (202 °C) than DMSO (189 °C),[70] the NMP – $PbI_2$ intermediate complex is not completely eliminated at the perovskite crystallization temperature (100 °C for 30 min)[71] and it therefore remains in the completely crystallized perovskite thin film. We intentionally annealed the perovskite film (obtained from 4:0:1) for 1 h and recorded the XRD diffractogram. As can be seen in **Figure S6**, the XRD peak intensity of the NMP – $PbI_2$ intermediate complex decreased when the perovskite film is annealed for 1h compared to the peak intensity of a perovskite film annealed for 30 min. A concomitant enhancement in the perovskite peak intensity and appearance of the $PbI_2$ peak is also observed. Although, longer annealing times can help to eliminate the NMP – $PbI_2$ intermediate complex and retain the residual $PbI_2$, it can also leads to the elimination of $MA^+$/$FA^+$ cations from the perovskite structure and the formation of defect/trap centers[71-74] and can be detrimental to the device performance. On the other hand, because of the weak van der Waals interaction between $PbI_2$ and NMP and the strong ionic interaction between $PbI_2$ and A-site cations,[54, 75-77] the NMP molecules intercalated in $PbI_2$ can be easily eliminated/replaced with strategies such as surface treatment of the perovskite with small or bulky organic ligands.[15, 54, 78, 79] The XRD results collectively suggest that with an optimal amount of NMP co-solvent a high-quality perovskite thin film can be obtained without any changes in the crystal structure compared to the perovskite layer obtained from the blends of traditional DMF:DMSO solvent mixture. On increasing the amount of NMP, the NMP – $PbI_2$ intermediate complex is formed which remains in the perovskite film and can be detrimental to the device performance. Scanning electron microscope (SEM) micrograph images, as shown in **Figure 3c** and **Figure 3d**, of perovskite films obtained from DMF:DMSO and DMF:DMSO:NMP (optimized ratio = 4:0.4:0.6) show no significant difference in the perovskite surface morphology. According to previous reports, the use of NMP co-solvent (e.g. in DMF:NMP) reduces the perovskite grain size via NMP – $PbI_2$ intermediate-to-perovskite phase transition.[51, 62] On the contrary, the similar perovskite grain size obtained from DMF:DMSO and DMF:DMSO:NMP (4:0.4:0.6) indicates that the perovskite is crystallized via the DMSO – $PbI_2$ intermediate complex route which is in line with the XRD results (no NMP-$PbI_2$ peak). This further indicates that the NMP is playing the sole role in enhancing the perovskite layer coverage on the hydrophobic Me-4PACz SAM. With an increase in the amount of NMP (in 4:0.2:0.8 and 4:0:1 (without DMSO) solvent ratios), a difference in the visual appearance of the perovskite film is observed (images are shown in Figure S5) indicating a non-uniform morphology. The SEM micrograph of perovskite films obtained from higher NMP co-solvent ratios (**Figure S7**) shows small grains with pinholes and a non-uniform morphology. In addition to this, small bright dots are observed on the perovskite surface (Figure S7) obtained from a higher concentration of NMP which can be assigned to NMP-$PbI_2$ crystalline intermediate complexes that are observed in the XRD pattern. True noncontact atomic force microscope (AFM) measurements (**Figure S8**) substantiate low root mean square (RMS) roughness's of 25 nm and 30 nm for perovskite film obtained from 4:0:0.4:0.6 and 4:0:1 (without DMSO) solvent ratio respectively and corroborates the top surface SEM morphology seen in Figures 3c and S7b.

To further understand the surface properties, high-resolution X-ray photoelectron spectroscopy (XPS) measurements on the surface of the perovskite layer are carried out. C1s and N1s core-level spectra of perovskite films obtained from 4:0.4:0.6 (**Figure 3e**), 4:0.2:0.8 (**Figure 3f**) and 4:0:1 (**Figure 3g**) solvent ratios shows chemical changes that occurred in the perovskite layer upon changes in the solvent ratio. In the XPS spectra of all three samples, the co-existence of FA and

MA components can be distinguished from each other by the C1s and N1s spectra. In the C1s spectra, the FA and MA components have specific peaks at ≈289 and ≈287 eV respectively. In the N1s spectra, the FA and MA peaks locate at ≈400.8 and ≈402.5 eV, respectively.[80] With increasing the NMP ratio, the peak intensity of MA (in C 1s and N 1s) is reduced while we observe no significant change in the peak intensity of FA. This indicates a loss of MA$^+$ cations with an increase of the NMP co-solvent ratio and changes in the overall perovskite composition. Previously, Lee et al., reported that the MA$^+$ cation (in MAPbI$_3$) has a stronger tendency to interact with DMSO and form a stable intermediate MAI-PbI$_2$-DMSO complex whereas NMP does not form a stable intermediate complex with MAI and PbI$_2$.[51] This indicates that when the amount of NMP increases (in 4:0.2:0.8 and 4:0:1 solvent ratios), the tendency to form a stable intermediate complex with MA$^+$ cation becomes less strong which might lead to evaporation of volatile MA$^+$ cations. This would explain the reduction in the MA peak intensity in the XPS spectra. Therefore, an optimal amount of triple co-solvent ratio (4:0.4:0.6) is vital to obtain a uniform perovskite film on hydrophobic Me-4PACz without any changes in the perovskite composition.

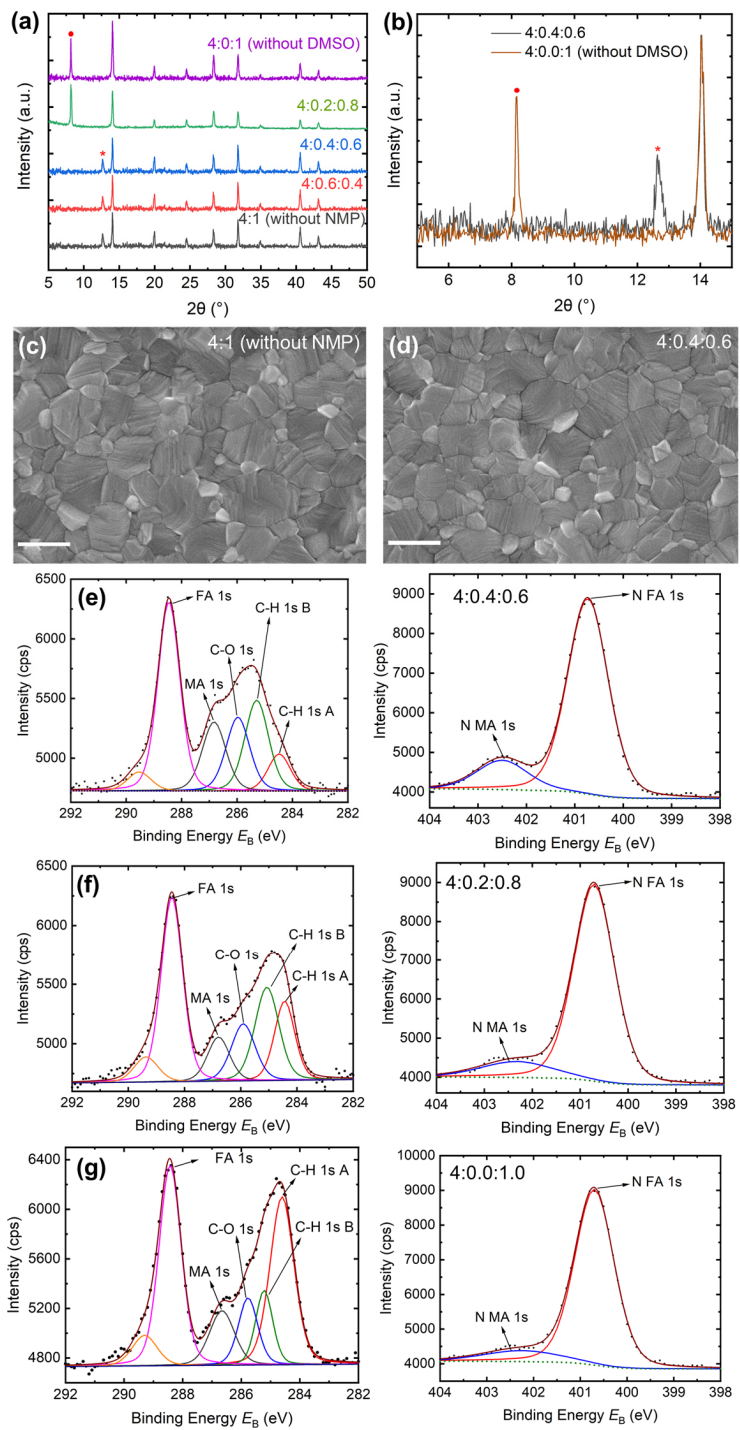

**Figure 3**: (a) XRD plot of perovskite thin films obtained from DMF:DMSO and triple co-solvent ratios (* = residual PbI$_2$ and • = PbI$_2$ – NMP complex), (b) XRD plot of perovskite films obtained from 4:0.4:0.6 and 4:0:1 (without DMSO) solvent ratios showing NMP-PbI$_2$ complex. SEM images of perovskite film obtained from (c) 4:1 (without NMP) and (d) with 4:0.4:0.6 solvent ratio. Scale bar, 200 nm. XPS of perovskite film obtained from DMF:DMSO:NMP with (e) 4:0.4:0.6, (f) 4:0.2:0.8 and (g) 4:0:1 solvent ratios.

*2.2.3 Device performance*

After the perovskite layer deposition on the Me-4PACz SAM, $C_{60}$ and BCP are evaporated as an electron transport layer (ETL) and a buffer layer, respectively, and the device is completed with Ag electrode evaporation. All details of the device fabrication are given in the Supporting Information. The *JV* curves of the best performing devices are shown in **Figure 4a** and the photovoltaic device parameters are tabulated in **Table S2**. It is important to note that the perovskite devices obtained from the DMF:DMSO solvent system (without NMP) are all short-circuited mainly because of incomplete perovskite layer coverage (Figure 1b) on the Me-4PACz SAM. With a change in solvent composition and an increase in the concentration of NMP, the photovoltaic device performance improved (Figure S5) alongside the perovskite layer coverage on Me-4PACz SAM. With a short-circuit current density ($J_{SC}$) of 21.74 mA/cm², a $V_{OC}$ of 1.15 V, and a fill factor (*FF*) of 80.73% and negligible hysteresis in the device *JV* curve compared to that of the other triple co-solvent ratio-based devices, the device obtained from the optimized triple co-solvent ratio of 4:0.4:0.6 demonstrated a champion efficiency of 20.18%. We note that the obtained efficiency is one of the highest for photovoltaic devices employing Me-4PACz HTL and a triple cation perovskite (with an $E_g$ = 1.63 eV). The improved efficiency is attributed to the formation of a uniform high-quality perovskite layer on Me-4PACz SAM and reduced interfacial traps at the perovskite/Me-4PACz interface.[31] **Figure S9** depicts the stabilized output under maximum power point tracking (MPP) reaching 19.6% (at 980 mV) which is in good agreement with the *JV* scan. The integrated photocurrent density calculated from the external quantum efficiency (EQE) spectrum (**Figure 4b**) agrees well with the $J_{SC}$ value measured from the *JV* curve. On the other hand, with an increase in the concentration of NMP in the triple solvent ratio, as expected, the photovoltaic device performance decreased. The *JV* curves and box plot of device parameters of devices obtained from higher concentration of NMP are shown in **Figure S10** and **Figure S11** respectively. Although with increasing the NMP ratio, the $V_{OC}$ and *FF* are slightly reduced, we attribute the main reason for the reduced photovoltaic performance to the reduction in $J_{SC}$ as shown in Figure S10. As discussed above, the formation and presence of NMP – $PbI_2$ crystalline intermediate complex in the perovskite film (XRD), pinholes, non-uniform perovskite morphology (SEM and AFM) and loss of $MA^+$ cations (XPS) from the perovskite structure accounts for the low perovskite solar cell efficiency.

To verify the reproducibility of the device performance with a triple co-solvent system, 18 devices (for NMP with a ratio of 0.2 and 0.4) and more than 75 devices for best optimized triple co-solvent ratio were fabricated using the device procedure outlined in Supporting Information. The distribution of photovoltaic parameters under reverse scan is summarized in **Figure S12a** – **Figure S12d**. The box in the statistical data contains 50% of the data points, the antenna shows minimum and maximum value, the point is the mean value, and the bar gives the median. The average performance improved from 7.1 ± 5 % (best value = 12.05%) for with 0.2 NMP ratio to ~19.5 ± 0.7 % (best value = 20.2%) for 0.6 NMP ratio with a narrow distribution range, where the enhancement is due to improved perovskite layer coverage and a concomitant improvement in $J_{SC}$, $V_{OC}$, and $FF$. This implies that with an optimal triple co-solvent system, we obtain highly reproducible perovskite solar cells with increased efficiencies. In addition to the role of NMP in improving the perovskite layer coverage on the Me-4PACz HTL, it becomes pertinent to verify whether the amount of NMP in the optimized triple co-solvents system is helping to improve the device performance. To verify this, we fabricated devices by depositing the perovskite layer on the MeO-2PACz HTL from blends of DMF:DMSO (without NMP) and the optimized DMF:DMSO:NMP solvent ratio. As illustrated in **Figure S13**, the best-performing device $JV$ curves and box plot distribution of device parameters showed a PCE of 21% and an average PCE of 19%, respectively, without significant differences. This indicates that the NMP present in the optimized triple co-solvent system plays the sole role in improving the interaction of the perovskite ink with the underlying Me-4PACz coated substrate and the perovskite layer coverage on the Me-4PACz HTL without compromising the device performance.

The poor performance reproducibility and lack of uniformity of the perovskite layer on hydrophobic HTLs make it challenging to obtain high efficiencies with larger area devices.[51, 62] Reproducible high efficiencies for a device with a cell area of 0.16 cm$^2$ motivated us to explore larger areas and we fabricated cells with 0.72 cm$^2$ and 1.08 cm$^2$ active area using our optimized perovskite ink-substrate interaction strategy, see **Figure 4c** and **Figure 4d** for the according $JV$ curves. The inset of Figure 4d shows the image of our large area ($A$ = 1.08 cm$^2$) perovskite device fabricated with our perovskite ink-substrate interaction strategy. The devices showed excellent performance with a PCE of 19.2 ± 0.3% (best value = 19.5%) and 17.7 ± 0.8% (best value = 18.5%) for 0.72 cm$^2$ and 1.08 cm$^2$ active area, respectively. **Figure 4e** and **Figure 4f** depicts the stabilized power output under maximum power point tracking reaching 19.5% (at 960 mV) and 18.05% (at

910 mV) for 0.72 cm$^2$ and 1.08 cm$^2$ active area devices, respectively, which are in good agreement with the *JV* scans. The device parameters of the best-performing device *JV* curves are tabulated in **Table S2** and the box plot of device parameters showing high reproducibility are shown in **Figure S14**. For both 0.72 cm$^2$ and 1.08 cm$^2$ cases, the cells showed negligible hysteresis in the *JV* curves as shown in **Figure S15a** and **Figure S15b**, respectively. Compared to the small cell area (0.16 cm$^2$), only ~ 4% and ~ 9% decrease in the PCE is observed in 0.72 cm$^2$ and 1.08 cm$^2$ device active area respectively, implying the importance of perovskite ink-substrate interaction strategy in obtaining a uniform perovskite film on hydrophobic Me-4PACz SAM.

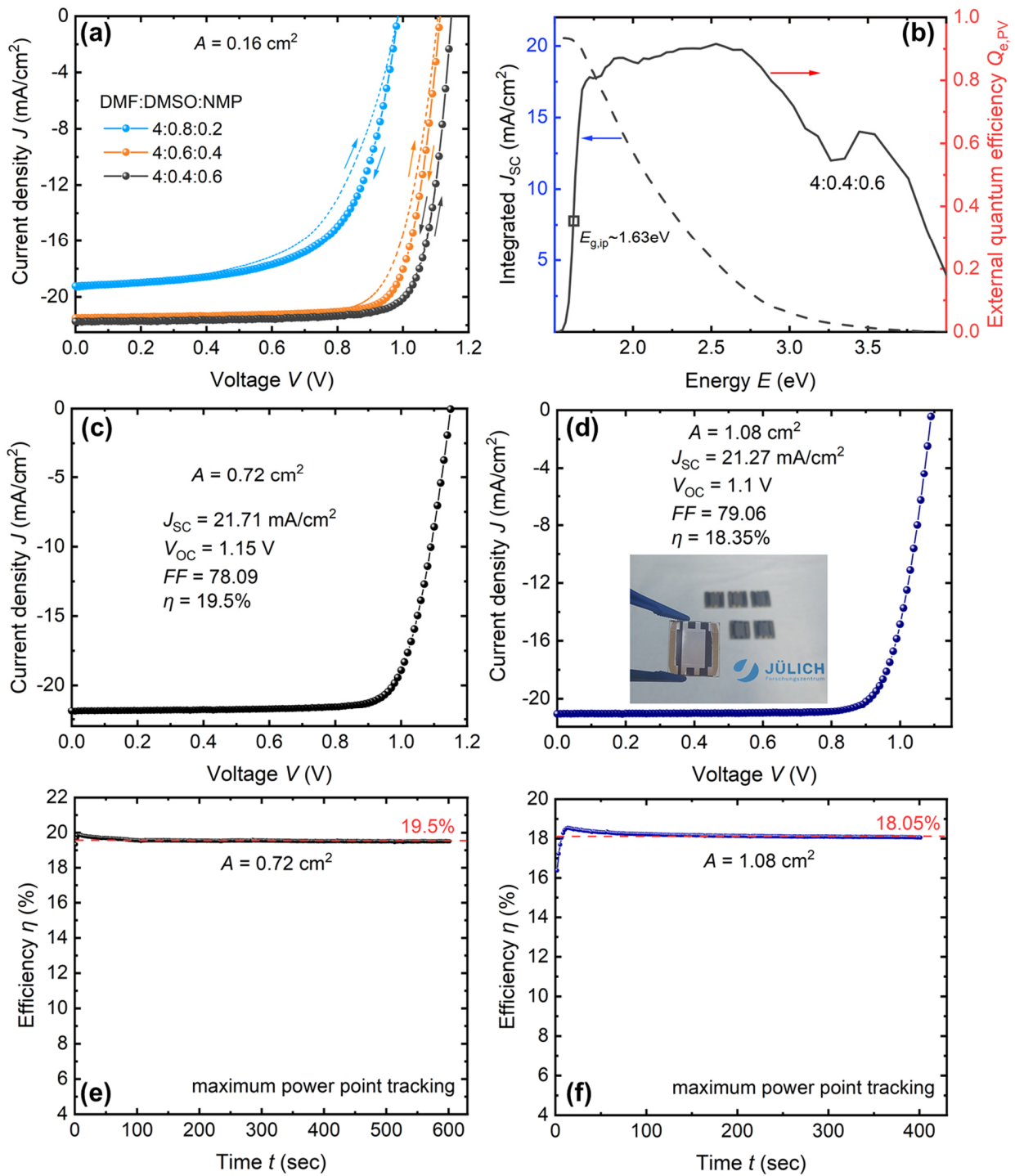

**Figure 4**: (a) Best performing *JV* curves of solar cells with different ratios of the DMF:DMSO:NMP co-solvent mixture measured under one Sun with forward and backward scan, (b) EQE spectrum of best performing device, and *JV* curves and of large area device with (c) 0.72 cm$^2$ active area and (d) 1.08 cm$^2$ active area. The inset shows the photographic image of our large area (1.08 cm$^2$) device. The power output under maximum power point tracking for (e) 0.72 cm$^2$

active area device with stabilized power output of 19.5% (at 960 mV) and (f) 1.08 cm² active area device with stabilized power output of 18.4% (at 910 mV). The voltage scan rate for all scans was 10 mV s⁻¹ and no device preconditioning, such as forward voltage bias applied for a long time or light soaking, was applied before starting the measurement.

### 2.2.4 Transient photoluminescence (tr-PL)

As discussed above, the perovskite devices obtained from different solvent ratios show different $V_{OC}$ and the solar cells with the optimized solvent ratio of 4:0.4:0.6 show the best $V_{OC}$ values as illustrated in **Figure 5a**. These differences in $V_{OC}$ refer to different recombination losses in the perovskite solar cells, and this should be reflected in the transient photoluminescence (tr-PL) decays.[81] To verify this in our samples, perovskite thin films of the four solvent ratios (DMF:DMSO:NMP = 4:1-x:x (0.4 > x > 1)) were measured using time correlated single photon counting technique. Each film was measured under three different laser fluences (814.87 nJ/cm², 80.98 nJ/cm², and 7.97 nJ/cm²). The background-corrected tr-PL decays were fitted using a rational function. We see the tr-PL decays with their fits versus time in **Figure 5b** for the four films of different solvent ratios under the lowest laser fluence (25 kHz using 2OD filter). At a certain laser fluence, the four films show different tr-PL decays and the film with the ratio of 4:0.4:0.6 shows the longest decay (the tr-PL decays of the four films under the other laser fluences can be found in the **Figure S16** in SI). As the PL decay scales exponentially with Fermi-level splitting (exp[$\Delta E_F$/kT]),[82] in **Figure 5c** we plot the differential decay time versus the Fermi-level splitting for the different solvent ratios under the different laser fluences. To obtain a decay time, the logarithms of the background-corrected decays were fitted using a rational function. These were then numerically differentiated to obtain a differential decay time according to

$$\tau_{\text{diff}}(t) = -2\left(\frac{d\ln(\phi(t))}{dt}\right)^{-1}$$

where $\phi(t)$ is the PL decay over time. The Fermi-level splitting $\Delta E_F$ is calculated according to

$$\Delta E_F = \Delta E_F(t=0) + kT\ln\left(\frac{\Phi(t)}{\Phi(t=0)}\right)$$

As can be seen from **Figure 5c**, the differential decay time is changing continuously with Fermi-level splitting for all the measured films and the sample with the optimized solvent ratio of 4:0.4:0.6 exhibits higher differential decay times compared to the other films. At lower Fermi-level splitting with the lowest laser fluence, the differential decay time goes to more than 10 μs for this ratio. In **Figure 5d**, we plot the average $V_{OC}$ of the measurements shown in Figure 5a for each of the different ratio devices versus the differential decay time at a constant Fermi-level splitting of 1.15 V. The deviation from the average $V_{OC}$ is also calculated and represented by the error bars for the four solvent ratio samples. At this constant Fermi-level splitting, the film with the solvent ratio of 4:0.4:0.6 shows the highest differential decay time of 5.95 μs and the highest average $V_{OC}$. The differential decay time at this constant Fermi-level splitting for the other ratios is 3.85 μs for 4:0.6:0.4 ratio, 2.26 μs for 4:0.2:0.8 ratio, and 1.53 μs for 4:0:1 ratio. The average $V_{OC}$ values are consistence with their corresponding differential decay time for all film ratios except for the sample with the ratio of 4:0.6:0.4 as it shows a wide variation of the measured $V_{OC}$.

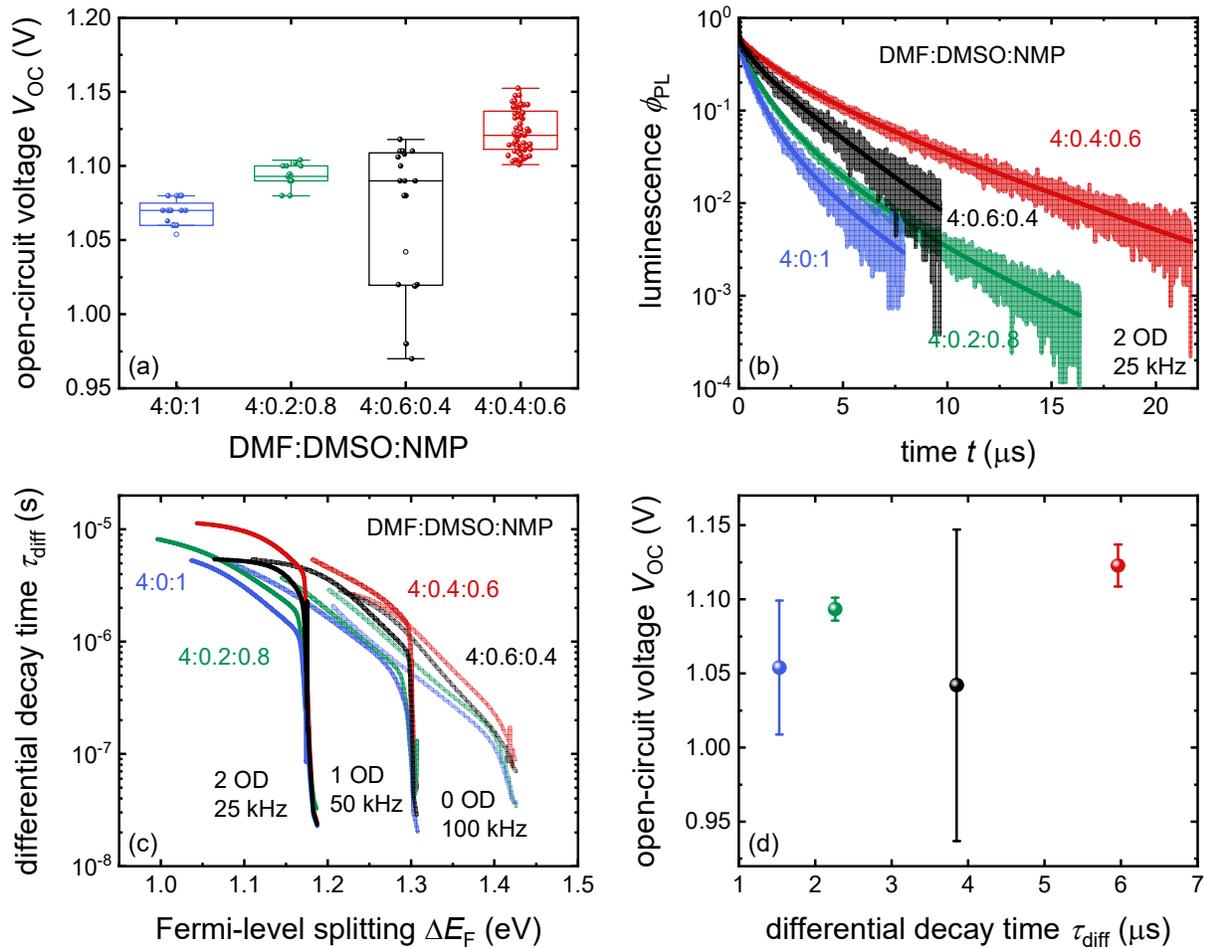

**Figure 5**: (a) The open circuit voltage $V_{OC}$ measured under 1 Sun for perovskite solar cells obtained from different solvent ratio of DMF:DMSO:NMP (4:0:1, 4:0.2:0.8, 4:0.6:0.4, and 4:0.4:0.6). (b)

Tr-PL decays of perovskite thin films of different solvent ratios of DMF:DMSO:NMP (4:0:1, 4:0.2:0.8, 4:0.6:0.4, and 4:0.4:0.6) measured with a laser frequency of 25 kHz using 2OD filter. The background-corrected decay data was fitted for each sample using a rational function (solid lines) (c) Differential decay times evaluated from the fitted tr-PL data as a function of Fermi-level splitting of all films of different solvent ratios measured with different laser frequencies and OD filters (25 kHz with 2OD, 50 kHz with 1OD, and 100 kHz with 0OD). (d) The average $V_{OC}$ represented by spheres for the four different solvent ratio devices in panel (a) versus the differential decay time at $\Delta E_F = 1.15$ V. The error bars represent the standard deviation from the average $V_{OC}$. The device with the solvent ratio of 4:0.4:0.6 shows the higher $V_{OC}$ and the film with this ratio shows the longer the tr-PL decay and the higher the differential decay time.

*2.2.5 Universality of triple co-solvent system*

Until recently, different perovskite compositions have been used by the community to study the photovoltaic performance, and therefore a universal method is of paramount significance for the development of the perovskite layer on the hydrophobic Me-4PACz SAM as HTL. Therefore, we use different perovskite compositions (MAPbI$_3$, FA$_{1-x}$MA$_x$PbI$_{3-y}$Br$_y$, MA-free FA$_{1-x}$Cs$_x$PbI$_{3-y}$Br$_y$, and all-inorganic CsPbI$_{1-x}$Br$_x$) to check the universality of our perovskite ink-substrate interaction strategy. These different perovskite compositions have been widely studied because of their high device performance and the application potential in silicon-perovskite and perovskite-perovskite tandem cells, semi-transparent photovoltaics, and indoor photovoltaics.[83-88] **Figure 6** shows the *JV* curves of devices employing MAPbI$_3$, FA$_{1-x}$MA$_x$PbI$_{3-y}$Br$_y$ and MA-free FA$_{1-x}$Cs$_x$PbI$_{3-y}$Br$_y$ perovskites obtained by improving the perovskite ink-substrate interaction on Me-4PACz SAM. The inset figure shows the photographic images of perovskite films obtained from DMF:DMSO and DMF:DMSO:NMP triple co-solvent system. Because of the incomplete perovskite layer formation with DMF:DMSO, the device performance is recorded only for the DMF:DMSO:NMP case. The device with MAPbI$_3$ showed 18 ± 1% (best value = 19%) (**Figure 6a**) with slight hysteresis (**Figure S18a**) in the *JV* curve, while devices employing FA$_{1-x}$MA$_x$PbI$_{3-y}$Br$_y$ and FA$_{1-x}$MA$_x$PbI$_{3-y}$Br$_y$ showed a hysteresis-free PCE of 19.5 ± 1.5% (best value = 21%) (**Figure 6b**) and 16 ± 2% (best value = 18%) (**Figure 6c**) respectively. The *JV* curves of FA$_{1-x}$MA$_x$PbI$_{3-y}$Br$_y$ and FA$_{1-x}$MA$_x$PbI$_{3-y}$Br$_y$ perovskite device showing no hysteresis is shown in **Figure S18b** and **Figure S18c** respectively. **Figure S19**, **Figure S20** and **Figure S21** show the box plots of the device parameters of MAPbI$_3$, FA$_{1-x}$MA$_x$PbI$_{3-y}$Br$_y$, and FA$_{1-x}$MA$_x$PbI$_{3-y}$Br$_y$ respectively, highlighting the reproducibility in the device performance with the perovskite ink-substrate interaction strategy. In addition to this, efforts were carried out to fabricate all-inorganic CsPbI$_2$Br devices by adding NMP as co-solvent. However, a high-temperature annealing step (180 °C) for inorganic perovskite crystallization[87] might not be suitable for underlying Me-4PACz SAM as all the devices were short-circuited. Nevertheless, we obtained uniform all-inorganic CsPbI$_2$Br perovskite thin films with the addition of NMP on Me-4PACz HTL. The images of inorganic perovskite thin films without and with NMP co-solvent strategy are shown in **Figure S22**.

These results evidence that a high-quality thin film of different perovskite compositions can be obtained on hydrophobic Me-4PACz by our perovskite ink-substrate interaction strategy, highlighting its universality.

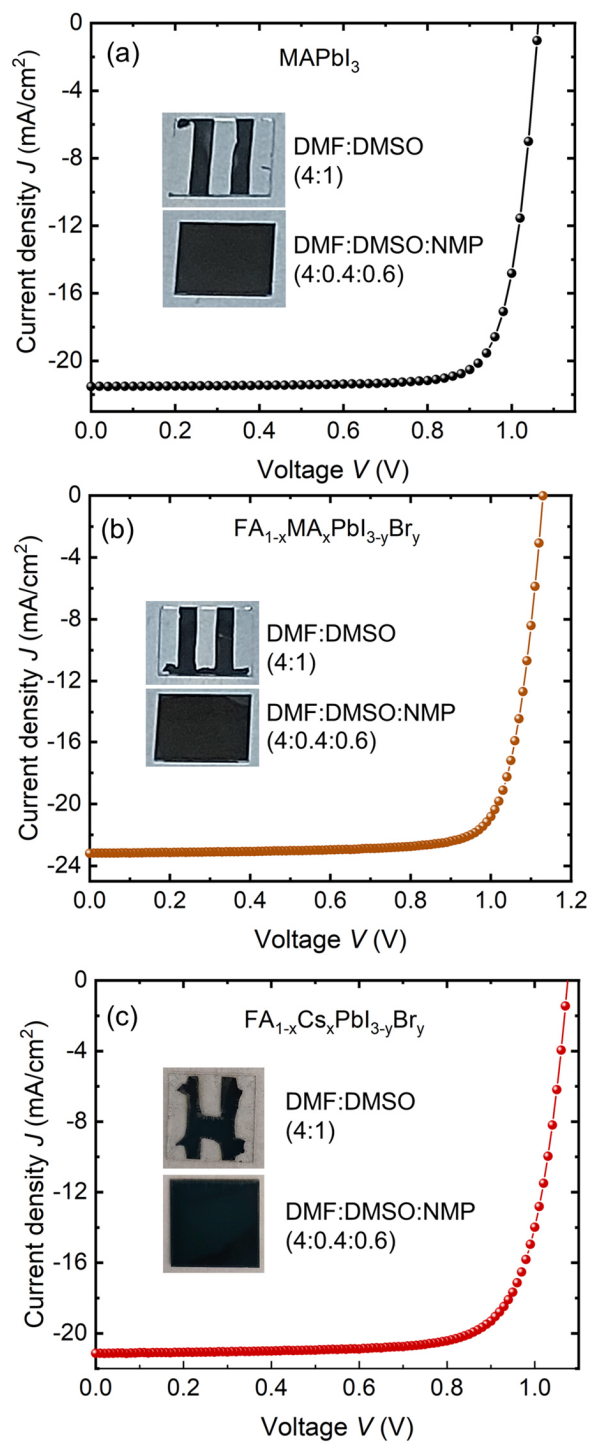

**Figure 6**: $JV$ curves of (a) $MAPbI_3$, (b) $FA_{1-x}MA_xPbI_{3-y}Br_y$ and (c) $FA_{1-x}Cs_xPbI_{3-y}Br_y$ obtained from perovskite ink-substrate interaction strategy on Me-4PACz SAM as HTL. The inset figure shows a photographic image of perovskite thin films obtained from DMF:DMSO and DMF:DMSO:NMP showing a uniform coverage in the latter case.

Lastly, we studied the long-term stability of triple cation perovskite solar cells from the triple co-solvent-based perovskite ink. The devices were stored and measured periodically over time under one Sun illumination intensity. The device PCEs as a function of storage time are summarized and shown in **Figure S23**. After around 3,600 h. of storage, the device retained 93% of the maximum device efficiency. Perovskite ink interaction with Me-4PACz coated substrate obtained from the blends of DMF:DMSO:NMP significantly aids not only in overcoming the hydrophobicity of Me-4PACz SAM but also in producing high-quality perovskite device with reproducible augmented efficiency and stability.

*2.3 Conclusion*

To summarize, we showed a perovskite ink-substrate interaction strategy comprising DMF:DMSO:NMP to obtain a uniform perovskite layer on a hydrophobic Me-4PACz SAM. Owing to its slightly non-polar nature, the NMP improved the binding of the perovskite ink with the Me-4PACz SAM, as elucidated by the binding energy calculations and contact angle measurements. In addition to this, it was found that the NMP co-solvent ratio (in DMF:DMSO:NMP) played a significant role in determining perovskite device performance. The optimized triple solvent ratio enabled high device efficiencies not only in a small area (0.16 cm$^2$) but also in devices with a large area (0.72 cm$^2$ and 1.08 cm$^2$) with high reproducibility. Moreover importantly, a uniform perovskite layer on Me-4PACz and high device efficiencies for MAPbI$_3$, FA$_{1-x}$MA$_x$PbI$_{3-y}$Br$_y$ and MA-free FA$_{1-x}$Cs$_x$PbI$_{3-y}$Br$_y$ cells were also achieved by this method, implying its universality. Hence, the problem of poor perovskite ink wetting on hydrophobic Me-4PACz can be well solved by our triple co-solvent system strategy. We believe that this perovskite ink-substrate interaction strategy can also be employed to obtain a uniform perovskite layer on other hydrophobic SAM and polymer HTLs.

**Supporting Information**

Supporting Information is available from the Wiley Online Library or from the author.

**Acknowledgements**

A.K. thanks the Helmholtz Young Investigator Group FRONTRUNNER. T.K. acknowledge funding from the Helmholtz Association via the project PEROSEED. M.S. thanks the German Research Foundation (DFG) for funding (SPP2196, GRK 2642). M.S. acknowledges funding by

ProperPhotoMile. Project ProperPhotoMile is supported under the umbrella of SOLAR-ERA.NET Cofund 2 by The Spanish Ministry of Science and Education and the AEI under the project PCI2020-112185 and CDTI project number IDI-20210171; the Federal Ministry for Economic Affairs and Energy on the basis of a decision by the German Bundestag project number FKZ 03EE1070B and FKZ 03EE1070A and the Israel Ministry of Energy with project number 220-11-031. SOLAR-ERA.NET is supported by the European Commission within the EU Framework Programme for Research and Innovation HORIZON 2020 (Cofund ERA-NET Action, No. 786483). Funded by the European Union. Views and opinions expressed are however those of the author(s) only and do not necessarily reflect those of the European Union or European Research Council Executive Agency (ERCEA). Neither the European Union nor the granting authority can be held responsible for them. The authors acknowledge funding from the European Research Council under the Horizon Europe programme (LOCAL-HEAT, grant agreement no. 101041809). R.S. and S.C. would like to acknowledge Center of Excellence in Materials and Manufacturing for Futuristic Mobility, Indian Institute of Technology (IIT) Madras for financial support and HRI Allahabad, and DST-SERB Funding (SRG/2020/001707) for the infrastructure. Computational work for this study was carried out at the cluster computing facility at IIT Madras. A.K. thanks Dr. Feray Ünlü and Prof. Sanjay Mathur for their help in performing contact angle measurements.

Received: ((will be filled in by the editorial staff))
Revised: ((will be filled in by the editorial staff))
Published online: ((will be filled in by the editorial staff))

## References

[1]     S. Sánchez, L. Pfeifer, N. Vlachopoulos, A. Hagfeldt, *Chemical Society Reviews* **2021**, *50*, 7108.

[2]     K. Liu, Y. Luo, Y. Jin, T. Liu, Y. Liang, L. Yang, P. Song, Z. Liu, C. Tian, L. Xie, Z. Wei, *Nature Communications* **2022,** *13*, 4891.

[3]     B. Chen, P. N. Rudd, S. Yang, Y. Yuan, J. Huang, *Chemical Society Reviews* **2019,** *48*, 3842.

[4]     Y. Xia, K. Sun, J. Chang, J. Ouyang, *Journal of Materials Chemistry A* **2015**, *3*, 15897.

[5]     N. Pant, A. Kulkarni, M. Yanagida, Y. Shirai, T. Miyasaka, K. Miyano, *Advanced Materials Interfaces* **2020,** *7*, 1901748.

[6]     C. M. Wolff, P. Caprioglio, M. Stolterfoht, D. Neher, *Advanced Materials* **2019,** *31*, 1902762.

[7]     S. Jariwala, H. Sun, G. W. P. Adhyaksa, A. Lof, L. A. Muscarella, B. Ehrler, E. C. Garnett, D. S. Ginger, *Joule* **2019,** *3*, 3048.

[8]     D. Bi, J. Luo, F. Zhang, A. Magrez, E. N. Athanasopoulou, A. Hagfeldt, M. Grätzel, *ChemSusChem* **2017,** *10*, 1624.

[9]     A. K. Jena, A. Kulkarni, T. Miyasaka, *Chemical Reviews* **2019,** *119*, 3036.

[10]    F. H. Isikgor, A. S. Subbiah, M. K. Eswaran, C. T. Howells, A. Babayigit, M. De Bastiani, E. Yengel, J. Liu, F. Furlan, G. T. Harrison, S. Zhumagali, J. I. Khan, F. Laquai, T. D. Anthopoulos, I. McCulloch, U. Schwingenschlögl, S. De Wolf, *Nano Energy* **2021,** *81*, 105633.

[11]    A. J. Doolin, R. G. Charles, C. S. P. De Castro, R. G. Rodriguez, E. V. Péan, R. Patidar, T. Dunlop, C. Charbonneau, T. Watson, M. L. Davies, *Green Chemistry* **2021,** *23*, 2471.

[12]    Y.-H. Seo, E.-C. Kim, S.-P. Cho, S.-S. Kim, S.-I. Na, *Applied Materials Today* **2017,** *9*, 598.

[13]    M. Kim, J. Jeong, H. Lu, T. K. Lee, F. T. Eickemeyer, Y. Liu, I. W. Choi, S. J. Choi, Y. Jo, H.-B. Kim, S.-I. Mo, Y.-K. Kim, H. Lee, N. G. An, S. Cho, W. R. Tress, S. M. Zakeeruddin, A. Hagfeldt, J. Y. Kim, M. Grätzel, D. S. Kim, *Science* **2022,** *375*, 302.

[14]    M. Saliba, T. Matsui, J.-Y. Seo, K. Domanski, J.-P. Correa-Baena, M. K. Nazeeruddin, S. M. Zakeeruddin, W. Tress, A. Abate, A. Hagfeldt, M. Grätzel, *Energy & Environmental Science* **2016,** *9*, 1989.

[15]    N. Pant, A. Kulkarni, M. Yanagida, Y. Shirai, S. Yashiro, M. Sumiya, T. Miyasaka, K. Miyano, *ACS Applied Energy Materials* **2021,** *4*, 4530.

[16]    C. C. Boyd, R. C. Shallcross, T. Moot, R. Kerner, L. Bertoluzzi, A. Onno, S. Kavadiya, C. Chosy, E. J. Wolf, J. Werner, J. A. Raiford, C. de Paula, A. F. Palmstrom, Z. J. Yu, J. J. Berry, S. F. Bent, Z. C. Holman, J. M. Luther, E. L. Ratcliff, N. R. Armstrong, M. D. McGehee, *Joule* **2020,** *4*, 1759.

[17]    A. T. Gidey, E. Assayehegn, J. Y. Kim, *ACS Applied Energy Materials* **2021,** *4*, 6923.

[18]    A. Kojima, K. Teshima, Y. Shirai, T. Miyasaka, *Journal of the American Chemical Society* **2009,** *131*, 6050.

[19] M. Degani, Q. An, M. Albaladejo-Siguan, Y. J. Hofstetter, C. Cho, F. Paulus, G. Grancini, Y. Vaynzof, *Science Advances* **2021,** *7*, eabj7930.

[20] D. Zhao, M. Sexton, H.-Y. Park, G. Baure, J. C. Nino, F. So, *Advanced Energy Materials* **2015,** *5*, 1401855.

[21] O. Malinkiewicz, A. Yella, Y. H. Lee, G. M. Espallargas, M. Graetzel, M. K. Nazeeruddin, H. J. Bolink, *Nature Photonics* **2014,** *8*, 128.

[22] C. Bi, Q. Wang, Y. Shao, Y. Yuan, Z. Xiao, J. Huang, *Nature Communications* **2015,** *6*, 7747.

[23] S. Chen, X. Xiao, B. Chen, L. L. Kelly, J. Zhao, Y. Lin, M. F. Toney, J. Huang, *Science Advances* **2021,** *7*, eabb2412.

[24] Z. Liu, L. Wang, C. Xu, X. Xie, Y. Zhang, *ACS Applied Energy Materials* **2021,** *4*, 10574.

[25] A. Al-Ashouri, A. Magomedov, M. Roß, M. Jošt, M. Talaikis, G. Chistiakova, T. Bertram, J. A. Márquez, E. Köhnen, E. Kasparavičius, S. Levcenco, L. Gil-Escrig, C. J. Hages, R. Schlatmann, B. Rech, T. Malinauskas, T. Unold, C. A. Kaufmann, L. Korte, G. Niaura, V. Getautis, S. Albrecht, *Energy & Environmental Science* **2019,** *12*, 3356.

[26] N. Phung, M. Verheijen, A. Todinova, K. Datta, M. Verhage, A. Al-Ashouri, H. Köbler, X. Li, A. Abate, S. Albrecht, M. Creatore, *ACS Applied Materials & Interfaces* **2022,** *14*, 2166.

[27] A. Al-Ashouri, E. Köhnen, B. Li, A. Magomedov, H. Hempel, P. Caprioglio, J. A. Márquez, A. B. Morales Vilches, E. Kasparavicius, J. A. Smith, N. Phung, D. Menzel, M. Grischek, L. Kegelmann, D. Skroblin, C. Gollwitzer, T. Malinauskas, M. Jošt, G. Matič, B. Rech, R. Schlatmann, M. Topič, L. Korte, A. Abate, B. Stannowski, D. Neher, M. Stolterfoht, T. Unold, V. Getautis, S. Albrecht, *Science* **2020,** *370*, 1300.

[28] D. B. Khadka, Y. Shirai, M. Yanagida, T. Tadano, K. Miyano, *Advanced Energy Materials* *n/a*, 2202029.

[29] X. Deng, F. Qi, F. Li, S. Wu, F. R. Lin, Z. Zhang, Z. Guan, Z. Yang, C.-S. Lee, A. K.-Y. Jen, *Angewandte Chemie International Edition* **2022,** *61*, e202203088.

[30] A. Al-Ashouri, E. Köhnen, B. Li, A. Magomedov, H. Hempel, P. Caprioglio, J. A. Márquez, A. B. M. Vilches, E. Kasparavicius, J. A. Smith, N. Phung, D. Menzel, M. Grischek, L. Kegelmann, D. Skroblin, C. Gollwitzer, T. Malinauskas, M. Jošt, G. Matič, B. Rech, R. Schlatmann, M. Topič, L. Korte, A. Abate, B. Stannowski, D. Neher, M. Stolterfoht, T. Unold, V. Getautis, S. Albrecht, *Science* **2020,** *370*, 1300.

[31] I. Levine, A. Al-Ashouri, A. Musiienko, H. Hempel, A. Magomedov, A. Drevilkauskaite, V. Getautis, D. Menzel, K. Hinrichs, T. Unold, S. Albrecht, T. Dittrich, *Joule* **2021,** *5*, 2915.

[32] K. Datta, J. Wang, D. Zhang, V. Zardetto, W. H. M. Remmerswaal, C. H. L. Weijtens, M. M. Wienk, R. A. J. Janssen, *Advanced Materials* **2022,** *34*, 2110053.

[33] A. Farag, T. Feeney, I. M. Hossain, F. Schackmar, P. Fassl, K. Küster, R. Bäuerle, M. A. Ruiz-Preciado, M. Hentschel, D. B. Ritzer, A. Diercks, Y. Li, B. A. Nejand, F. Laufer, R. Singh, U. Starke, U. W. Paetzold, *Advanced Energy Materials n/a*, 2203982.

[34] M. Taddei, J. A. Smith, B. M. Gallant, S. Zhou, R. J. Westbrook, Y. Shi, J. Wang, J. N. Drysdale, D. P. McCarthy, S. Barlow, *ACS Energy Letters* **2022,** *7*, 4265.

[35] P. Tockhorn, J. Sutter, A. Cruz, P. Wagner, K. Jäger, D. Yoo, F. Lang, M. Grischek, B. Li, J. Li, O. Shargaieva, E. Unger, A. Al-Ashouri, E. Köhnen, M. Stolterfoht, D. Neher, R. Schlatmann, B. Rech, B. Stannowski, S. Albrecht, C. Becker, *Nature Nanotechnology* **2022,** *17*, 1214.

[36] L. Li, Y. Wang, X. Wang, R. Lin, X. Luo, Z. Liu, K. Zhou, S. Xiong, Q. Bao, G. Chen, Y. Tian, Y. Deng, K. Xiao, J. Wu, M. I. Saidaminov, H. Lin, C.-Q. Ma, Z. Zhao, Y. Wu, L. Zhang, H. Tan, *Nature Energy* **2022,** *7*, 708.

[37] R. Mishima, M. Hino, M. Kanematsu, K. Kishimoto, H. Ishibashi, K. Konishi, S. Okamoto, T. Irie, T. Fujimoto, W. Yoshida, H. Uzu, D. Adachi, K. Yamamoto, *Applied Physics Express* **2022,** *15*, 076503.

[38] M. Roß, S. Severin, M. B. Stutz, P. Wagner, H. Köbler, M. Favin-Lévêque, A. Al-Ashouri, P. Korb, P. Tockhorn, A. Abate, B. Stannowski, B. Rech, S. Albrecht, *Advanced Energy Materials* **2021,** *11*, 2101460.

[39] M. Roß, L. Gil-Escrig, A. Al-Ashouri, P. Tockhorn, M. Jošt, B. Rech, S. Albrecht, *ACS Applied Materials & Interfaces* **2020,** *12*, 39261.

[40] A. Al-Ashouri, M. Marčinskas, E. Kasparavičius, T. Malinauskas, A. Palmstrom, V. Getautis, S. Albrecht, M. D. McGehee, A. Magomedov, *ACS Energy Letters* **2023,** 898.

[41] C.-H. Kuan, G.-S. Luo, S. Narra, S. Maity, H. Hiramatsu, Y.-W. Tsai, J.-M. Lin, C.-H. Hou, J.-J. Shyue, E. Wei-Guang Diau, *Chemical Engineering Journal* **2022,** *450*, 138037.

[42] N. Tzoganakis, B. Feng, M. Loizos, M. Krassas, D. Tsikritzis, X. Zhuang, E. Kymakis, *Journal of Materials Chemistry C* **2021,** *9*, 14709.

[43] T. Leijtens, T. Giovenzana, S. N. Habisreutinger, J. S. Tinkham, N. K. Noel, B. A. Kamino, G. Sadoughi, A. Sellinger, H. J. Snaith, *ACS Applied Materials & Interfaces* **2016,** *8*, 5981.

[44] H. O. S. Yadav, A.-T. Kuo, S. Urata, W. Shinoda, *The Journal of Physical Chemistry C* **2020,** *124*, 14237.

[45] R. Godawat, S. N. Jamadagni, S. Garde, *Proceedings of the National Academy of Sciences* **2009,** *106*, 15119.

[46] Y. Gao, L. Duan, S. Guan, G. Gao, Y. Cheng, X. Ren, Y. Wang, *RSC Advances* **2017,** *7*, 44673.

[47] W. Chen, V. Karde, T. N. H. Cheng, S. S. Ramli, J. Y. Y. Heng, *Frontiers of Chemical Science and Engineering* **2021,** *15*, 90.

[48] B. Chaudhary, A. Kulkarni, A. K. Jena, M. Ikegami, Y. Udagawa, H. Kunugita, K. Ema, T. Miyasaka, *ChemSusChem* **2017,** *10*, 2473.

[49] J.-W. Lee, H.-S. Kim, N.-G. Park, *Accounts of Chemical Research* **2016,** *49*, 311.

[50] S. Wang, A. Wang, X. Deng, L. Xie, A. Xiao, C. Li, Y. Xiang, T. Li, L. Ding, F. Hao, *Journal of Materials Chemistry A* **2020,** *8*, 12201.

[51] J.-W. Lee, Z. Dai, C. Lee, H. M. Lee, T.-H. Han, N. De Marco, O. Lin, C. S. Choi, B. Dunn, J. Koh, D. Di Carlo, J. H. Ko, H. D. Maynard, Y. Yang, *Journal of the American Chemical Society* **2018,** *140*, 6317.

[52] S. N. Manjunatha, Y.-X. Chu, M.-J. Jeng, L.-B. Chang, *Journal of Electronic Materials* **2020,** *49*, 6823.

[53] F. Cataldo, *European Chemical Bulletin* **2015,** *4*, 92.

[54] W. S. Yang, J. H. Noh, N. J. Jeon, Y. C. Kim, S. Ryu, J. Seo, S. I. Seok, *Science* **2015,** *348*, 1234.

[55] S. Casalini, C. A. Bortolotti, F. Leonardi, F. Biscarini, *Chemical Society Reviews* **2017,** *46*, 40.

[56] P. Xue, P. Wang, P. Chen, B. Yao, P. Gong, J. Sun, Z. Zhang, R. Lu, *Chemical Science* **2017,** *8*, 6060.

[57] R.-F. Dou, X.-C. Ma, L. Xi, H. L. Yip, K. Y. Wong, W. M. Lau, J.-F. Jia, Q.-K. Xue, W.-S. Yang, H. Ma, A. K. Y. Jen, *Langmuir* **2006,** *22*, 3049.

[58] F. M. Hoffmann, *Surface Science Reports* **1983,** *3*, 107.

[59]     H.-Y. Nie, M. J. Walzak, N. S. McIntyre, *The Journal of Physical Chemistry B* **2006,** *110*, 21101.

[60]     K. Nakamura, T. Takahashi, T. Hosomi, Y. Yamaguchi, W. Tanaka, J. Liu, M. Kanai, K. Nagashima, T. Yanagida, *ACS Omega* **2022,** *7*, 1462.

[61]     L. R. Snyder, *Journal of Chromatographic Science* **1978,** *16*, 223.

[62]     K. O. Brinkmann, J. He, F. Schubert, J. Malerczyk, C. Kreusel, F. van gen Hassend, S. Weber, J. Song, J. Qu, T. Riedl, *ACS Applied Materials & Interfaces* **2019,** *11*, 40172.

[63]     C. Griffin W, *J. Soc. Cosmet. Chem.* **1954,** *5*, 249.

[64]     C. Griffin W, *J. Soc. Cosmet. Chem.* **1949,** *1*, 311.

[65]     W. C. Griffin, presented at **1954**.

[66]     W. C. Griffin, presented at **1946**.

[67]     Y. Jo, K. S. Oh, M. Kim, K.-H. Kim, H. Lee, C.-W. Lee, D. S. Kim, *Advanced Materials Interfaces* **2016,** *3*, 1500768.

[68]     F. Cheng, X. Jing, R. Chen, J. Cao, J. Yan, Y. Wu, X. Huang, B. Wu, N. Zheng, *Inorganic Chemistry Frontiers* **2019,** *6*, 2458.

[69]     S. J. Lee, J. H. Heo, S. H. Im, *ACS Applied Materials & Interfaces* **2020,** *12*, 8233.

[70]     J. C. Hamill, Jr., J. Schwartz, Y.-L. Loo, *ACS Energy Letters* **2018,** *3*, 92.

[71]     M. Wang, C. Fei, M. A. Uddin, J. Huang, *Science Advances* **2022,** *8*, eabo5977.

[72]     C. Zhang, A. Baktash, J.-X. Zhong, W. Chen, Y. Bai, M. Hao, P. Chen, D. He, S. Ding, J. A. Steele, T. Lin, M. Lyu, X. Wen, W.-Q. Wu, L. Wang, *Advanced Functional Materials* **2022,** *32*, 2208077.

[73]     J. Thiesbrummel, V. M. Le Corre, F. Peña-Camargo, L. Perdigón-Toro, F. Lang, F. Yang, M. Grischek, E. Gutierrez-Partida, J. Warby, M. D. Farrar, S. Mahesh, P. Caprioglio, S. Albrecht, D. Neher, H. J. Snaith, M. Stolterfoht, *Advanced Energy Materials* **2021,** *11*, 2101447.

[74]     R. D. J. Oliver, P. Caprioglio, F. Peña-Camargo, L. R. V. Buizza, F. Zu, A. J. Ramadan, S. G. Motti, S. Mahesh, M. M. McCarthy, J. H. Warby, Y.-H. Lin, N. Koch, S. Albrecht, L. M. Herz, M. B. Johnston, D. Neher, M. Stolterfoht, H. J. Snaith, *Energy & Environmental Science* **2022,** *15*, 714.

[75]     M. Hiroshi, N. Yoshinaho, N. Miharu, *Chemistry Letters* **1980,** *9*, 663.

[76]     N. J. Jeon, J. H. Noh, Y. C. Kim, W. S. Yang, S. Ryu, S. I. Seok, *Nature Materials* **2014,** *13*, 897.

[77]	C. Wu, X. Zheng, Q. Yang, Y. Yan, M. Sanghadasa, S. Priya, *The Journal of Physical Chemistry C* **2016,** *120*, 26710.

[78]	N. Pant, A. Kulkarni, M. Yanagida, Y. Shirai, T. Miyasaka, K. Miyano, *ACS Applied Energy Materials* **2020,** *3*, 6215.

[79]	Q. Jiang, Y. Zhao, X. Zhang, X. Yang, Y. Chen, Z. Chu, Q. Ye, X. Li, Z. Yin, J. You, *Nature Photonics* **2019,** *13*, 460.

[80]	T. J. Jacobsson, J.-P. Correa-Baena, E. Halvani Anaraki, B. Philippe, S. D. Stranks, M. E. F. Bouduban, W. Tress, K. Schenk, J. Teuscher, J.-E. Moser, H. Rensmo, A. Hagfeldt, *Journal of the American Chemical Society* **2016,** *138*, 10331.

[81]	L. Krückemeier, B. Krogmeier, Z. Liu, U. Rau, T. Kirchartz, *Advanced Energy Materials* **2021,** *11*, 2003489.

[82]	T. Kirchartz, J. A. Márquez, M. Stolterfoht, T. Unold, *Advanced Energy Materials* **2020,** *10*, 1904134.

[83]	P. Wu, D. Thrithamarassery Gangadharan, M. I. Saidaminov, H. Tan, *ACS Central Science* **2022**.

[84]	Y. Deng, S. Xu, S. Chen, X. Xiao, J. Zhao, J. Huang, *Nature Energy* **2021,** *6*, 633.

[85]	Y. Wang, T. Mahmoudi, Y.-B. Hahn, *Advanced Energy Materials* **2020,** *10*, 2000967.

[86]	A. K. Jena, A. Kulkarni, Y. Sanehira, M. Ikegami, T. Miyasaka, *Chemistry of Materials* **2018,** *30*, 6668.

[87]	S. Öz, A. K. Jena, A. Kulkarni, K. Mouri, T. Yokoyama, I. Takei, F. Ünlü, S. Mathur, T. Miyasaka, *ACS Energy Letters* **2020,** *5*, 1292.

[88]	B. Parida, S. Yoon, J. Ryu, S. Hayase, S. M. Jeong, D.-W. Kang, *ACS Applied Materials & Interfaces* **2020,** *12*, 22958.